# Mechanism of unidirectional movement of kinesin motors

Ping Xie, Shuo-Xing Dou, and Peng-Ye Wang

*Laboratory of Soft Matter Physics, Institute of Physics, Chinese Academy of Sciences,*

*P .O. Box, 603, Beijing 100080, China*

## Abstract

Kinesin motors have been studied extensively both experimentally and theoretically. However, the microscopic mechanism of the processive movement of kinesin is still an open question. In this paper, we propose a hand-over-hand model for the processivity of kinesin, which is based on chemical, mechanical, and electrical couplings. In the model the processive movement does not need to rely on the two heads' coordination in their ATP hydrolysis and mechanical cycles. Rather, the ATP hydrolyses at the two heads are independent. The much higher ATPase rate at the trailing head than the leading head makes the motor walk processively in a natural way, with one ATP being hydrolyzed per step. The model is consistent with the structural study of kinesin and the measured pathway of the kinesin ATPase. Using the model the estimated driving force of ~ 5.8 pN is in agreements with the experimental results (5~7.5 pN). The prediction of the moving time in one step (~10 $\mu$s) is also consistent with the measured values of 0~50 $\mu$s. The previous observation of substeps within the 8-nm step is explained. The shapes of velocity-load (both positive and negative) curves show resemblance to previous experimental results.

*Keywords*:   kinesin; molecular motor; processivity; mechanism

Conventional kinesin is a two-headed molecular motor protein that converts chemical energy of ATP hydrolysis to mechanical force. It can translocate processively and unidirectionally along a microtubule for hundreds of 8-nm steps without dissociating (1–5). A single kinesin molecule can exert maximal forces of 5–7.5 pN (3, 6–9), and at low load its velocity can reach 800 nm/s (5, 9). Conventional kinesin transports membrane-bound vesicles and organelles in various cells (10–12).

Since its discovery, it has been studied extensively by using different experimental methods, such as biochemical, biophysical, X-ray crystallography and cryo-EM, many aspects of its movement behavior have been gradually elucidated. In particular, the rapid development and progress of single-molecule manipulation and detection techniques in recent years (13, 14) have improved significantly our knowledge of its dynamic and mechanistic properties *in vitro*, with important parameters such as stall force, velocity, mean run length, and dwell time being determined systematically (3–9, 15–21).

The microscopic mechanism of the processive movement of kinesin is still not very clear. Based on experimental results, several models have been proposed. One is the thermal ratchet model in which a motor is viewed as a Brownian particle moving in two (or more) periodic but spatially asymmetric stochastically switched potentials (22–24). Another is the "hand-over-hand" model (1, 22, 25–32). In this model, it is supposed that the motor maintains continuous attachment to the microtubule by alternately repeating single-headed and double-headed binding. Adjacent tubulin heterodimmers on the microtubule serve as the consecutive binding sites. This model requires that the two heads move in a coordinated manner and alternately move past each other. This coordination is realized by a mechanical communication between the two heads during their ATP hydrolysis cycles. The third model postulates that kinesin head movement is coordinated through an "inchworm" mechanism (22, 33–36), in which also at least one head remains bound to the microtubule during the kinesin movement but the two heads do not swap places as different from the case in the symmetric hand-over-hand model.

For quantitatively studying the motion of kinesin, multistate chemical kinetic description is often used (18, 37, 38). In this approach, it is postulated that the motor protein molecule steps through a sequence of discrete chemical states linked by rate constants. This approach relies mainly on biochemical observations and data. It can explain well some experimental results on the mechanical behaviors of kinesin, such as velocity, mean run length, and their load dependence.

In this paper, we present a hand-over-hand model that relies on chemical, mechanical, and electrical couplings. In this model, the processive movement does not require the coordination between the two heads. The dimeric kinesin steps forward very naturally and, in general, one ATP is hydrolyzed per step (1:1 coupling). This movement of kinesin with 1:1 coupling results solely from that the ATPase rate at the trailing head is much higher than the leading head, which is caused by the different forces acting on the two heads (the trailing head being pulled forward and the leading head backward). Using the model we study the kinesin dynamics quantitatively and give good explanations to some experimental results which have not been well explained previously.

**Structural Consideration**

The structural study of dimeric kinesin protein, for example, the rat brain kinesin, by using X-ray crystallography reveals that one kinesin head has +8 net elementary charges (35). By using X-ray crystallography and cryo-electron microscopy, Kikkawa *et al*. (39) reveal that each tubulin heterodimer of microtubules, i.e., the α-tubulin and β-tubulin monomers, has –27 net elementary charges (the α-tubulin monomer having –12 net charges and the β-tubulin monomer having –15 net charges). Thus we assume that the interaction between kinesin heads and tubulin heterodimers is electrostatic. When a kinesin head is not very close to a tubulin heterodimer, the electrostatic force can be approximated as being produced by two ideal charged particles. When the kinesin head is close enough to the tubulin heterodimer, the electrostatic force is dependent on the charge distributions on the two surfaces (40, 41). So that the kinesin head will bind the tubulin heterodimer in a fixed orientation (28, 42–44).

From X-ray crystallography and cyro-EM observations, it is revealed that the two kinesin heads in rigor states bind two successive tubulin heterodimers of microtubule in equivalent orientations (28, 42, 43), as schematically shown in Fig. 1 (a).

The determination of the crystal structure of kinesin dimer by X-ray crystallography (35) shows that its equilibrium state (or free state) corresponds to a structure as schematically shown in Fig. 1(b). The two heads are related by a 120º rotation about an axis close to that of the coiled-coil neck (the direction of which is perpendicular to the paper surface), with an equilibrium distance (center-of-mass distance) as ~5 nm. Marx *et al*. (45) further reveal that the structure of kinesin dimer in solution is similar to the crystal structure, with the center-of-mass distance between the two heads being slightly greater. It is taken for granted that this equilibrium state corresponds to the state that the kinesin dimer has the minimum free energy. Thus the state shown in Fig. 1(a) corresponds to a state that the kinesin dimer has a greater free energy since it has a very large conformational change from its equilibrium state both in center-of-mass distance and relative orientation of the two heads.

Based on the principle of minimum free energy, it is anticipated that once one of the two heads or both become(s) free, the kinesin dimer structure tends to change from the state in Fig. 1(a) to that in Fig. 1(b). This change is realized mainly via an elastic force and an elastic torque between the two heads.

**Model**

Basing on the above analysis we propose a physical model to describe the processive movement of two-headed kinesin motors moving unidirectionally along microtubules. We begin with the two heads of kinesin binding to two successive tubulin heterodimers (separated by 8 nm) of the microtubules by electrostatic forces between the positively-charged heads and negatively-charged tubulin heterodimers, as seen in Fig. 1(a). In this rigor state, the structure of the kinesin dimer is highly strained. In accordance with the principle of minimum free energy, there will exist an elastic force and an elastic torque exerting on the two heads to induce the kinesin dimer to change from the state in Fig. 1(a) to that in Fig. 1(b), besides an electrostatic force exerted by the negatively-charged tubulin heterodimer.

Activated by the microtubule, the ATPase cycles at the two heads begin. We consider two cases. (i) The ATP molecule bound to the trailing head is hydrolyzed first. The chemical reaction of the hydrolysis of ATP to ADP and $P_i$ results in the weakening of the binding electrostatic force (46). One of our assumptions on the weakening of electrostatic force is as follows. The chemical reaction leads to the change of electrical property of the local solution, for example, causing the relative dielectric constant, $\varepsilon_r = 1 + \chi_e$, to increase. This may be due to an increase of shielding effect on the negative charges of the tubulin heterodimer, which is in turn induced by the increasing number of positive ions, for example, $H^+$, $K^+$ and $Mg^{2+}$ in the solution, congregating in the local solution where the chemical reaction has just taken place. This increase of relative dielectric constant of the local solution induces the electrostatic force to become smaller than the elastic force. Thus the trailing head becomes "free" (free from binding to microtubule), and the kinesin dimer changes from the state in Fig. 1(a) to that in Fig. 1(b). In the equilibrium state of Fig. 1(b) the elastic force is zero, and there exists only the electrostatic force exerted by the neighboring tubulin heterodimer (III) of the microtubule. This electrostatic force induces binding of the free head to the tubulin heterodimer (III) as in Fig. 1(c). A forward step of 8 nm is made with one ATP being hydrolyzed (1:1 coupling). The two heads exchange their relative positions and roles in the succeeding mechanical cycle. (ii) The ATP molecule bound to the leading head is hydrolyzed first. The leading head then becomes "free". Thus kinesin dimer changes from the state in Fig. 1(a') to the equilibrium state in Fig. 1(b'). When the original electrical property of the local solution near the tubulin heterodimer (II) is recovered, for example, after the congregated positive ions diffuse away, the leading head will bind to the tubulin heterodimer (II) again by the recovered strong electrostatic force. Then the ATP molecule bound to the trailing head hydrolyzes and the kinesin makes a forward step as in case (i). Two ATP molecules are hydrolyzed to make this forward step (2:1 coupling). After this step the ATP molecule bound to the "new" trailing head has a larger probability to hydrolyze earlier than the "new" leading head if the ATPase rates at the two heads are assumed to be the same.

For ATPase rates (including both ATP binding and turnover rates) at the two heads, we consider two cases. (i) The ATPase rates at the two heads are the same. In this case, the ATP molecules bound to the trailing and leading heads have equal probability to be hydrolyzed first at the beginning. Therefore, for making the first forward step there will be equal probabilities of hydrolyzing one ATP (1:1 coupling) and of hydrolyzing two ATP (2:1 coupling). After the first forward step is made, there will be a higher probability of hydrolyzing one ATP molecule than of hydrolyzing two ATP per step. (ii) In the rigor state as shown in Fig. 1(a), there exists a forward (i.e., the plus-end directed) force on the neck linker of the trailing head and a backward force on that of the leading head. As will be seen below, according to the energy-landscape model the ATPase rate at the trailing head will be enhanced whereas that at the leading head be reduced. Thus even at the beginning, the probability is high that ATP is hydrolyzed earlier at the trailing head than at the leading head. Therefore, the kinesin motor generally hydrolyzes one ATP molecule per step (1:1 coupling).

**Results and Discussion**

**Driving Force.** We give a simplest estimation of the magnitude of driving force for the forward movement of kinesin basing on this model. Before the equilibrium state as shown in Fig. 1(b) is reached, the force exerting on the trailing head is mainly the elastic force. As the equilibrium state is reached, only an electrostatic force exerts on the free head by the neighboring tubulin heterodimer (III). For simplicity, if we consider the tubulin heterodimer as a particle with $q_1 = -27$ charges and the kinesin head as a particle with $q_2 = +8$ charges, then the electrostatic force will be $F = \frac{1}{4\pi\varepsilon_r\varepsilon_0}\frac{q_1 q_2}{r^2}$, where the value of the relative dielectric constant for the solution is taken to be $\varepsilon_r = 78$ (40). For simplicity, we approximate the ellipsoidal kinesin head as a sphere with radius $r_k \approx 3 \text{ nm}$. Referring to Fig. 2, the distance between the positive charge center of the free kinesin head and the negative charge center of tubulin heterodimer (III) is $r = \sqrt{\left[d - r_0 \cos(\pi/6)\right]^2 + \left[r_0 \sin(\pi/6) + d_{vertical}\right]^2} = 7.04$ nm, where $d$ = 8 nm is the distance between the two successive tubulin heterodimers, and $r_0 = 5.5 \text{ nm}$ is the equilibrium distance between the two heads of kinesin dimer in solution. Considering the fact that the negative charges are mainly distributed on the surfaces of the tubulin heterodimers, we take the vertical distance between the center of the bound kinesin head and the negative charge center of tubulin heterodimer as $d_{vertical} = 3.5 \text{ nm}$. Thus the electrostatic force is $F \approx 12.8 \text{ pN}$. Its component in the forward direction is ~ 5.8 pN, which is the driving force at this head position. This is consistent with the measured stall force of 5.5 ~ 7.5 pN (3, 6–9), which is the one necessary to stop the motion of kinesin. Detailed discussion on stall force will be given later.

**Leap Time.** We estimate the leap time $\tau$ of the moving head which is defined as the time for the head to move from tubulin heterodimer (I) to heterodimer (III) in Fig. 1(a). To this end, we resort to the following equation for an over-damped Brownian particle

$$\Gamma\, dx/dt = F_{driving} - F_{load} + f(t),  \qquad [1]$$

where $\Gamma$ is the frictional drag coefficient, $v = dx/dt$ is the moving velocity of the head along the microtubule, $F_{driving}$ is the driving force on the head, and $F_{load}$ is the force exerted on the head by load. $f(t)$ is the fluctuating Langevin force, with $\langle f(t) \rangle = 0$ and $\langle f(t)f(t') \rangle = 2k_B T \Gamma \delta(t-t')$. From the Stokes formula we obtain $\Gamma = 6\pi\eta r_k = 5.65 \times 10^{-11} \text{kg s}^{-1}$, where the viscosity $\eta$ of the aqueous medium of a cell around kinesin is approximately $0.01 \text{ g cm}^{-1} \text{ s}^{-1}$. For simplicity, we approximate the driving force as a constant at any position of the moving head between tubulin heterodimers (I) and (III) and is equal to that when the head is at the equilibrium position as in Fig. 1(b), i.e., $F_{driving} = 5.8 \text{ pN}$. In experiments the positive load is exerted on kinesin in the opposite direction of its movement with a bead connected to the two heads through a coiled-coil neck and neck linkers. Thus it can be approximately considered that the load is always exerted on the *head at leading position*. So that the measured rise time of the bead movement by Nishiyama *et al.* (20) (or the moving time of the bead in one step) is the time for the free head to make its second half step (corresponding to positions *A* and *C* of the free head, shown in dashed blues lines in Fig. 2), i.e., the rise time is about half of the leap time $\tau$ of the free head. When there is no load the position of the kinesin dimer is its center-of-mass position and thus the moving time of kinesin in one step is the leap time $\tau$. The load is $F_{load} = F_{trap} + 6\pi\eta v r_{bead}$, where $F_{trap}$ is the force by the optical trap and $6\pi\eta v r_{bead}$ is the Stokes force of the bead, with $r_{bead}$ being the radius of the bead. Taking the experimental value of $r_{bead} = 0.2 \text{ μm}$ and the optical trap force $> 3$ pN (20), from Eq. **1** we obtain the rise time as $5.4 \text{ μs}$ at $F_{trap} = 3 \text{ pN}$, $8.5 \text{ μs}$ at $F_{trap} = 4 \text{ pN}$, and $19 \text{ μs}$ at $F_{trap} = 5 \text{ pN}$. These values of rise time are consistent with the measured values of $0 \sim 50 \text{ μs}$ (20). Note that these values of rise time are much smaller than that of the ATP turnover time, which is in the order of 10 ms (5, 9, 18). When there is no load we obtain the leap time as $\tau = 156 \text{ ns}$.

In fact, the second half step as mentioned above is composed of two substeps: One is from position *A* to equilibrium position *B*, during which the head is mainly driven by a forward elastic force. The other is from *B* to *C*, during which the head is driven by a backward elastic force and a forward electrostatic force. As a result, the net forward driving force during the second substep may become smaller than the first substep. Thus the first substep is fast and the second substep is slow. The size of the first substep from $A$ to $B$ is ~4.7 nm, and that of the second substep is ~3.3 nm. These are qualitatively consistent with the observed substeps within the 8-nm step of single kinesin molecules (20).

The energy required to make a forward step is $E \approx 5.8 \text{ pN} \times 16 \text{ nm} = 92.8 \times 10^{-21} \text{ J}$. This is consistent with the free energy released from ATP hydrolysis $(\sim 25 k_B T = 104 \times 10^{-21} \text{ J})$.

**Dynamics.** We give a discussion of the moving velocity $V$ of kinesin versus ATP concentration [ATP] and load $F_{load}$ basing on the model. As the moving time of kinesin in one step is about three orders smaller than the ATPase time, the two heads are almost always bound to the microtubule during a whole ATPase cycle and $V$ is actually only dependent on the ATPase rates at the two heads. The ATPase rate at each head should satisfy the Michaelis-Menten kinetics

$$K = \frac{k_c[\text{ATP}]}{[\text{ATP}] + k_c/k_b},  \quad [2]$$

where the ATP turnover rate $k_c$ and ATP binding rate $k_b$ follow the general Boltzman form (47)

$$k_c = \frac{k_c^{(0)}(1 + A_c)}{1 + A_c \exp(F\delta/k_B T)}, \quad k_b = \frac{k_b^{(0)}(1 + A_b)}{1 + A_b \exp(F\delta/k_B T)}. \quad [3]$$

The forces $F$ on the neck linkers of the two heads can be written as

$$F^T = F_{load} - F_0, \quad [4a]$$

$$F^L = F_{load} + F_0, \quad [4b]$$

where $F_0$ is the internal elastic force, $F^T$ and $F^L$ are for the trailing and leading heads, respectively. Note that $F_{load} = F_{trap}$ here because the motor is in its rigor state.

Since for a positive load $F_{load} \geq 0$ the ATPase rate at the trailing head is much faster ( > 500 times) than that at the leading head (which will be seen below), $V$ is essentially only dependent on the ATPase rate $K^T$ at the trailing head, *i.e.*, $V = K^T \times 8 \text{ nm}$. Using Eqs. **2-4a**, the experimentally measured velocity of kinesin versus [ATP] and load (Figs. 2 and 3a in ref. 9) can be fitted very well. The theoretical curves are shown in Fig. 3. Using the parameter values in Fig. 3 we calculate the ATPase rates at the two heads for $F_{load} = 0$, which are shown in Fig. 4. It can be seen that the ATPase rate at the trailing head is several orders larger than that at the leading head.

When $F_{load}$ becomes negative (*i.e.*, forward load), the ATPase rate at the leading head will increase more significantly and may become comparable to that at the trailing head. Thus we must consider the contributions of both heads to kinesin movement. First we consider that ATP binding rate at the leading head, $k_b^L$, is not yet close to that at the trailing head, $k_b^T$. This corresponds to the case that the load is positive or it is negative but not large enough. In this case, the probability of ATP hydrolysis at the leading head is still negligible, but that of ATP binding cannot be neglected. As the leading head will become a new trailing head in the successive mechanical cycle, the contribution of $k_b^L$ to the movement of kinesin is equivalent to increasing the ATP binding rate at the trailing head. The effective ATP binding rate at the trailing head can be written as (see Appendix A)

$$k_{b-eff}^T = \frac{k_b^T + k_b^L}{1 - k_b^L/k_c^T}, \qquad [5]$$

whereas the ATP turnover rate $k_{c-eff}^T$ is still $k_c^T$. From Eqs. **2-5** we give a result for velocity *V* as a function of load (both positive and negative) in Fig. 5. At low [ATP], ATP binding is rate limiting and thus *V* is mainly determined by $k_{b-eff}^T$, which increases at most by 2-fold from 0 pN to high negative load, i.e., $k_{b-eff}^T \approx k_b^T + k_b^L \approx 2k_b^T$ (because $k_c^T \gg k_b^L$ at low [ATP]). This corresponds to the case in Figs. 5(b) and (c) where the saturating *V* at high negative load is about two times that at 0 pN. At high [ATP], ATP turnover is rate limiting and thus *V* is mainly determined by $k_c^T$, which is already saturated when the load decrease from positive to 0 pN. Thus *V* increases only slightly with negative load. This corresponds to the case in Fig. 5(a).

The shape of the load-velocity curves in Fig. 5 shows good resemblance to the experimental curves in Fig. 5 of ref. 8, especially when [ATP] is low. However, the measured *V* has a 3-fold increase from 0 pN to high negative load at low [ATP] [Figs. 5*b* and *c* in ref. 8] and ~ 50% increase at high [ATP] (Fig. 5*a* in ref. 8). We explain this discrepancy as follows. In the calculations, we have neglected the contribution of ATP hydrolysis at the leading head in the previous cycle to the ATP turnover after the leading head becomes the trailing head in the current cycle. That is, we use $k^T_{c-eff} = k^T_c$ in the calculations. In fact, when the ATP binding rate at the leading head, $k^L_b$, is close to that at the trailing head, $k^T_b$, the probability is high that ATP hydrolysis at the leading head in the previous cycle has begun before this head becomes the trailing head in the current cycle. This is equivalent to reducing the ATP turnover time (i.e., increasing the ATP turnover rate) at the trailing head, which results in the increase of ATPase rate $K^T$ at the trailing head even at saturating [ATP]. From the experimental result at saturating [ATP] (Fig. 5*a* in ref. 8), we may deduce that the effective ATP turnover rate $k^T_{c-eff}$ has a ~50% increase at high negative load over that at no load. We expect that, at low [ATP], the ATPase rate (or *V*) will also have the same ~50% increase due to the increase of $k^T_{c-eff}$. Therefore, due to the increases of both $k^T_{b-eff}$ and $k^T_{c-eff}$ the velocity *V* at low [ATP] is totally increased by ~3-fold at high negative load over that at no load. For a quantitative calculation of $k^T_{c-eff}$, the probability distributions of ATP binding and turnover rates instead of their mean values at the two heads are required to consider.

When the forward load is increased further, the probability becomes high that ATP hydrolysis is started and finished earlier at the leading head than at the trailing head. Thus the probability of two ATPase cycles being coupled to one forward step for kinesin becomes high, resulting in a decrease of the moving velocity. This explains qualitatively why the moving velocity starts to drop off when the forward load is further increased in the experiment (8).

**Stall Force.** Finally, we give an explanation of the [ATP] dependence of stall force as measured by Visscher *et al*. (9). As we mentioned above, if there were no noise the stall force would be a constant value of 5.8 pN, *i.e.*, the driving force. When the noise is present, the stall force, *i.e.*, the minimum force to stop forward movement of the free trailing head, should be equal to the driving force $F_{driving}$ plus the "diffusion force" $F_{diff}$ required to stop the free head diffusing $2d \approx 16$ nm in the forward direction. The diffusion of the free head with force $F_{diff}$ satisfies equation

$$dx/dt = -f_{diff} + \tilde{f}(t),  \quad\quad\quad [6]$$

where $f_{diff} = F_{diff}/\Gamma$. $\tilde{f}(t)$ has the property $\langle \tilde{f}(t) \rangle = 0$, $\langle \tilde{f}(t)\tilde{f}(t') \rangle = 2D\delta(t-t')$,

where $D = k_B T/\Gamma$ is the Einstein diffusion constant. From the above equation the mean first-passage time $T$ at which the free head diffuse a distance of 2*d* has the form (see Appendix B)

$$T = \frac{1}{f_{diff}} \left[ \frac{D}{f_{diff}} \left( \exp(\frac{f_{diff}}{D} 2d) - 1 \right) - 2d \right], \quad \text{[7]}$$

where the mean first-passage time $T$ should be equal to the free time of the trailing head. The free time is determined by the recovery time of the electrical property of the local environment of the solution near the tubulin heterodimer (I) in Fig. 1(a) and is assumed to be in the same order of the ATP turnover time. It is taken as $t_{free} \approx$ 10 ms. When there are multiple ATPase cycles for the trailing head, the total free time is approximately $(K^T t_M) t_{free}$ and thus we have

$$T = K^T t_M t_{free}, \quad \text{[8]}$$

where $t_M = 2$ s is the time required to measure the stall force in the experiment (9) and $K_T$ is the ATPase rate at the trailing head with the load $F_{trap}$ being equal to the stall force $F_{stall} = F_{driving} + F_{diff}$. Using parameter values in Fig. 3, $D = k_B T/6\pi\eta r_k$, Eqs. **2-4a**, **7** and **8**, we calculate $F_{stall}$ versus [ATP]. The result is shown in Fig. 6. It is seen that the theoretical curve shows similar behavior to the experimental result (Fig. 3b of ref. 9). Compared with the experimental result, the theoretical value is slightly higher. This may be due to that we use the mean first-passage time instead of the first-passage time distribution and that $(K^T t_M) t_{free}$ is an overestimate of the total free time.

In summary, basing on the previous structural and kinetic studies of dimeric kinesin motors we present a chemically-, mechanically- and electrically-coupled model to describe the unidirectional movement of kinesin. Using the model the estimated driving force is consistent with previous experimental results. The theoretical predictions of the moving time (*i.e.*, rise time) in one step and existence of substeps, the velocity versus [ATP] and loads (both positive and negative), and the dependence of the stall force on [ATP] are in agreement with previous experiments.

This research is supported by the National Natural Science Foundation of China (Grant number: 60025516).

**Appendix A**

We start with the moment when the trailing head just leaps forward and becomes the leading head as in Fig. 1(a). Assume the dwell time of kinesin between adjacent steps is $t_{dwell}$ and neglect the leap time of the trailing head in each step, then $t_{dwell}$ is the time for the leading head to wait for ATP binding. It is also the ATPase time for the trailing head, thus we have

$$t_{dwell} = \tau_{b-eff}^{T} + \tau_{c}^{T}, \quad [A1]$$

where $\tau_{c}^{T} = 1/k_{c}^{T}$ and $\tau_{b-eff}^{T}$ is the time taken by the trailing head for ATP binding and thus is $\tau_{b-eff}^{T} = 1/k_{b-eff}^{T}$. As the trailing head has spent time $t_{dwell}$ in the previous cycle and thus has a ATP binding probability $k_{b}^{L} t_{dwell}$, thus $\tau_{b-eff}^{T}$ should satisfy

$$k_{b}^{L} t_{dwell} + k_{b}^{T} \tau_{b-eff}^{T} = 1. \quad [A2]$$

From Eqs. **A1** and **A2**, Eq. **5** is obtained. It should be noted that Eq. **5** is only an approximation. The precise calculation of $k_{b-eff}^{T}$ need to consider the probability distributions of ATP binding and turnover rates at the two heads.

**Appendix B**

From Langevin equation (6) we have the following Fokker-Planck equation

$$\frac{\partial W(x,t)}{\partial t} = f_{diff} \frac{\partial W(x,t)}{\partial x} + D \frac{\partial^2 W(x,t)}{\partial^2 x}.$$

Using Eq. (5.2.157) in ref. 48, from the above equation we have

$$\psi(x) = \exp\left[\int_{a}^{x} (-\frac{f_{diff}}{D}) dx\right] = \exp\left[-\frac{f_{diff}}{D}(x-a)\right],$$

where the motion is supposed in the interval ($a$, $b$) and the barrier at $a$ is reflecting and the barrier at $b$ is absorbing. In our case $2d = b - a$. Using Eq. (5.2.160) in ref. 48, the mean first-passage time can be written as

$$T(a) = 2\int_{a}^{b} \frac{dy}{\psi(y)} \int_{a}^{y} \frac{\psi(z)}{2D} dz.$$

Integrating the above equation we obtain Eq. **7**.

**References**

1. Howard, J., Hudspeth, A. J. & Vale R. D. (1989) Movement of microtubules by single kinesin molecules. *Nature* **342,** 154–158.
2. Block, S. M., Goldstein, L. S. & Schnapp, B. J. (1990) Bead movement by single kinesin molecules studied with optical tweezers. *Nature* **348**, 348–352.
3. Svoboda, K., Schmidt, C. F., Schnapp, B. J. & Block, S. M. (1993) Direct observation of kinesin stepping by optical trapping interferometry. *Nature* **365,** 721–727.
4. Schnitzer, M. J. & Block, S. M. (1997) Kinesin hydrolyzes one ATP per 8-nm step. *Nature* **388**, 386–390.
5. Hua, W., Young, E. C., Fleming, M. L. & Gelles, J. (1997) Coupling of kinesin steps to ATP hydrolysis. *Nature* **388,** 390–393.
6. Svoboda, K. & Block, S. M. (1994) Force and velocity measured for single kinesin molecules. *Cell* **77**, 773–784.
7. Meyhöfer, E. & Howard, J. (1995) The force generated by a single kinesin molecule against an elastic load. *Proc. Natl. Acad. Sci. USA* **92**, 574–578.
8. Coppin, C. M., Pierce, D.W., Hsu, L. & Vale, R. D. (1997) The load dependence of kinesin's mechanical cycle. *Proc. Natl. Acad. Sci. USA* **94,** 8539–8544.
9. Visscher, K., Schnitzer, M. J. & Block, S. M. (1999) Single kinesin molecules studied with a molecular force clamp. *Nature* **400,** 184–189.
10. Vale, R. D. & Fletterick, R. J. (1997) The design plan of kinesin motors. *Annu. Rev. Cell Dev. Biol.* **13,** 745–777.
11. Hirokawa, N. (1998) Kinesin and dynein superfamily proteins and the mechanism of organelle transport. *Science* **279**, 519–526.
12. Vale, R. D. (2003) The molecular motor moolbox for intracellular transport. *Cell* **112,** 467–480.
13. Mehta, A. D., Rief, M., Spudich, J. A., Smith, D. A. & Simmons, R. M. (1999) Single-molecule biomechanics with optical methods. *Science* **283,** 1689–1695.
14. Ishii, Y., Ishijima, A. & Yanagida, T. (2001) Single molecule nanomanipulation of biomolecules. *Trends Biotech.* **19,** 211–216.
15. Coppin, C. M., Finer, J. T., Spudich, J. A. & Vale, R. D. (1996) Detection of sub-8-nm movements of kinesin by high-resolution optical-trap microscopy. *Proc. Natl. Acad. Sci. USA* **93,** 1913–1917.

16. Vale, R. D., Funatsu, T., Pierce, D. W., Romberg, L., Harada, Y. & Yanagida, T. (1996) Direct observation of single kinesin molecules moving along microtubules. *Nature* **380,** 451–453.

17. Kojima, H., Muto, E., Higuchi, H. & Yanagida, T. (1997) Mechanics of single kinesin molecules measured by optical trapping nanometry. *Biophys. J.* **73,** 2012–2022.

18. Schnitzer, M. J., Visscher, K. & Block, S. M. (2000) Force production by single kinesin motors. *Nat. Cell Biol.* **2,** 718–723.

19. Yajima, J., Alonso, M. C., Cross, R. A. & Toyoshima, Y. Y. (2002) Direct Long-Term Observation of Kinesin Processivity at Low Load. *Curr. Biol.* **12,** 301–306.

20. Nishiyama, M., Muto, E., Inoue, Y., Yanagida, T. & Higuchi, H. (2001) Substeps within the 8-nm step of the ATPase cycle of single kinesin molecules. *Nat. Cell Biol.* **3,** 425–428.

21. Nishiyama, M., Higuchi, H. & Yanagida, T. (2002) Chemomechanical coupling of the forward and backward steps of single kinesin molecules. *Nat. Cell Biol.* **4,** 790–797.

22. Endow, S. A. & Barker, D. S. (2003) Processive and nonprocessive models of kinesin movement. *Annu. Rev. Physiol.* **65,** 161–175.

23. Jülicher, F., Ajdari, A. & Prost, J. (1997) Modeling molecular motors. *Rev. Mod. Phys.* **69,** 1269–1281.

24. Astumian, R. D. & Derenyi, I. (1999) A chemically reversible Brownian motor: application to kinesin and Ncd. *Biophys. J.* **77**, 993–1002.

25. Schnapp, B. J., Crise, B., Sheetz, M. P., Reese, T. S. & Khan, S. (1990) Delayed start-up of kinesin-driven microtubule gliding following inhibition by adenosine 59-[b,g-imido]triphosphate. *Proc. Natl. Acad. Sci. USA* **87,** 10053–10057.

26. Hackney, D. D. (1994) Evidence for alternating head catalysis by kinesin during microtubule-stimulated ATP hydrolysis. *Proc. Natl. Acad. Sci. USA* **91**, 6865–6869.

27. Howard, J. (1996) The Movement of Kinesin Along Microtubules. *Annu. Rev. Physiol.* **58**, 703–729.

28. Mandelkow, E. & Johnson, K. A. (1998) The structural and mechanochemical cycle of kinesin. *Trends Biochem. Sci.* **23,** 429–433.

29. Rice, S., Lin, A. W., Safer, D., Hartß, C. L., Naberk, N., Carragher, B. O., Cain, S.

M., Pechatnikova, E., Wilson-Kubalek, E. M., Whittaker, M., PateI, E., Cookek, R., Taylor, E. W., Milligan, R. A. & Vale, R. D. (1999) A structural change in the kinesin motor protein that drives motility. *Nature* **402,** 778–784.

30. Hancock, W. O. & Howard, J. (1999) Kinesin's processivity results from mechanical and chemical coordination between the ATP hydrolysis cycles of the two motor domains. *Proc. Natl. Acad. Sci. USA* **96**, 13147–52.

31. Vale, R. D. & Milligan, R. A. (2000) The way things move: looking under the hood of molecular motor proteins. *Science* **288,** 88–95.

32. Kawaguchi K. & Ishiwata, S. (2001) Nucleotide-dependent single to double-headed binding of kinesin. *Science* **291,** 667–669.

33. Block, S. M. & Svoboda, K. (1995) Analysis of high resolution recordings of motor movement. *Biophys. J.* **68,** 230s–241s.

34. Lockhart, A., Crevel, I. M. & Cross, R. A. (1995) Kinesin and ncd bind through a single head to microtubules and compete for a shared MT binding site. *J. Mol. Biol.* **249,** 763–771.

35. Kozielski, F., Sack, S., Marx, A., Thormahlen, M., Schonbrunn, E., Biou, V., Thompson, A., Mandelkow, E.-M. & Mandelkow, E. (1997) The crystal structure of dimeric kinesin and implications for microtubule-dependent motility. *Cell* **91,** 985–994.

36. Hua, W., Chung, J. & Gelles, J. (2002) Distinguishing inchworm and hand-over-hand processive kinesin movement by neck rotation measurements. *Science* **295,** 844–848.

37. Keller, D. & Bustamante, C. (2000) The mechanochemistry of molecular motors. *Biophys. J.* **78,** 541–556.

38. Fisher, M. E. & Kolomeisky, A. B. (2001) Simple mechanochemistry describes the dynamics of kinesin molecules. *Proc. Natl. Acad. Sci. USA* **98,** 7748–7753.

39. Kikkawa, M., Sablin, E. P., Okada, Y., Yajima, H., Fletterick, R. J. & Hirokawa, N. (2001) Switch-based mechanism of kinesin motors. *Nature* **411,** 439-445.

40. Honig, B. & Nicholls, A. (1995) Classical electrostatics in biology and chemistry. *Science* **268,** 1144–1149.

41. Vale, R. D. (1999) Millennial musings on molecular motors. *Trends Cell Biol.* **9,** M38–M42.

42. Hoenger, A., Sack, S., Thormahlen, M., Marx, A., Muller, J., Gross, H. & Mandelkow, E. (1998) Image reconstruction of microtubules decorated with monomeric and dimeric kinesins: comparison with X-ray structure and implications for motility. *J. Cell Biol.* **141,** 419–430.

43. Mandelkow, E. & Hoenger, A. (1999) Structures of kinesin and kinesin–microtubule interactions. *Curr. Opin. Cell Biol.* **11,** 34–44.
44. Hirose, K., Löwe, J., Alonso, M., Cross, R. A. & Amos, L. A. (1999) Congruent docking of dimeric kinesin and ncd into three-dimensional electron cryomicroscopy maps of microtubule–motor ADP complexes. *Mol. Biol. Cell*, **10,** 2063–2074.
45. Marx, A., Thormählen, M., Müller, J., Sack, S., Mandelkow, E.-M. & Mandelkow, E. (1998) Conformations of kinesin: solution vs. crystal structures and interactions with microtubules. *Eur. Biophys. J.* **27,** 455– 465.
46. Hancock,W. O. & Howard, J. (1999) Kinesin's processivity results from mechanical and chemical coordination between the ATP hydrolysis cycles of the two motor domains. *Proc. Natl. Acad. Sci. USA* **96,** 13147–13152.
47. Wang, M. D., Schnitzer, M. J., Yin, H., Landick, R., Gelles, J. & Block, S. M. (1998) Force and velocity measured for single molecules of RNA polymerase. *Science* **282,** 902–907.
48. Gardiner, C.W. (1983) *Handbook of Stochastic Methods for Physics, Chemistry and the Natural Sciences* (Springer-Verlag, Berlin).

**Figure captions**

**Fig. 1.** Schematic illustrations of kinesin movement mechanism. The two kinesin heads are in blue and red. The *α*- and *β*-tubulin subunits from a single microtubule protofilament are indicated in light green and light blue. Each pair of *α*- and *β*-tubulin subunits forms a tubulin heterodimer. The polarity of the microtubule is indicated. **Effective mechanochemical cycle:** (a) The cycle begins with both heads binding to the microtubule. Note that the elastic force exerts on the two heads in opposite directions. (b) ATP hydrolysis at the blue head (trailing head) changes the electrical property of the local solution, causing an increase of the dielectric constant of the local solution. This in turn weakens the electrostatic force between this head and the tubulin heterodimer (I), resulting in the detachment and subsequent movement of the blue head towards a position as determined by the equilibrium structure of the kinesin dimer. (c) The blue head binds to the new tubulin heterodimer (III) via the electrostatic interactions and becomes the leading head for the next cycle. One ATP is hydrolyzed for this 8-nm forward step. **Futile mechanochemical cycle:** (a') The cycle also begins with both heads binding to the microtubule. (b') ATP hydrolysis at the red head (leading head) weakens the electrostatic force between this head and the tubulin heterodimer (II), resulting in the detachment and subsequent movement of the red head towards the equilibrium position. (c') The red head rebinds to the tubulin heterodimer (II) after the original electrical property of local environment of the solution is recovered. One ATP is hydrolyzed in this futile mechanical cycle.

**Fig. 2.** Geometry used to demonstrate the substeps and to calculate the distance *r* between the positive charge center of the free kinesin head and the negative charge center of tubulin heterodimer. Position *A* corresponds to that the centre of mass of the free head has the same horizontal coordinate as the bound head in red, and position *B* corresponds to the equilibrium position of the free head. *d* is the period of the microtubule protofilament, $r_0$ is the free distance between the two heads of kinesin dimer in solution, and $d_{vertical}$ is the vertical distance between the center of the bound kinesin head and the negative charge center of tubulin heterodimer.

**Fig. 3.** (a) Velocity versus [ATP] at various loads. (b) Load - velocity curves at [ATP] = 5 μM and [ATP] = 2 mM. The parameter values: $\delta = 3.7$ nm, $k_c^{(0)} = 49.1$ s$^{-1}$, $A_c = 1.11$, $k_b^{(0)} = 0.16$ μM$^{-1}$s$^{-1}$, $A_b = 7.43$. The elastic force between the two heads is taken as $F_0 = 5.8$ pN.

**Fig. 4.** ATPase rates at the trailing head and leading head with no load. The parameter values are the same as in Fig. 3.

**Fig. 5.** Load - velocity curves at (a) [ATP] = 5 μM, (b) [ATP] = 40 μM, and (c) [ATP] = 1 mM. The parameter values: $F_0 = 6$ pN, $\delta = 3.7$ nm, $k_c^{(0)} = 49.1$ s$^{-1}$, $A_c = 1.11$, $k_b^{(0)} = 0.16$ μM$^{-1}$s$^{-1}$, $A_b = 0.62$.

**Fig. 6.** Stall force versus [ATP]. $F_{driving} = 5.8$ pN and other parameter values are the same as in Fig. 3.

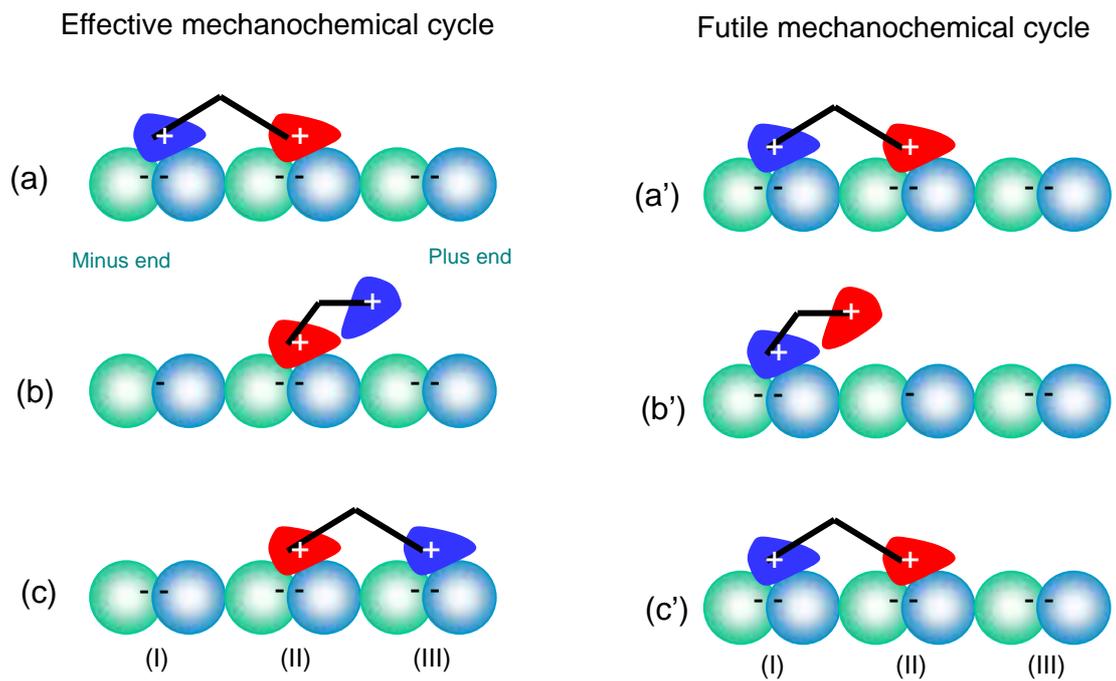

Fig. 1

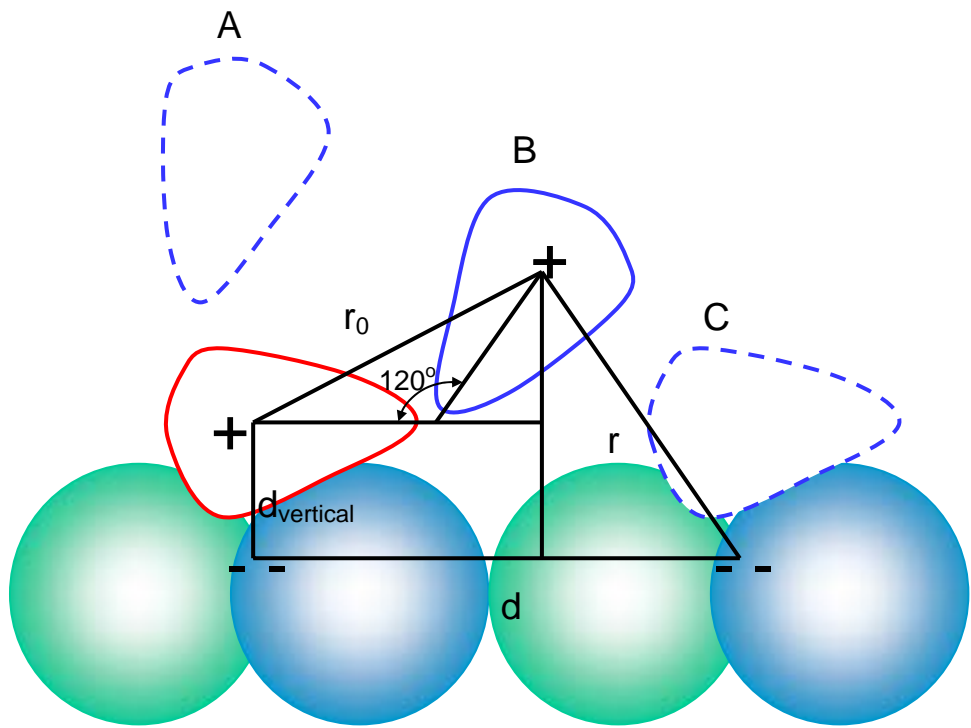

Fig. 2

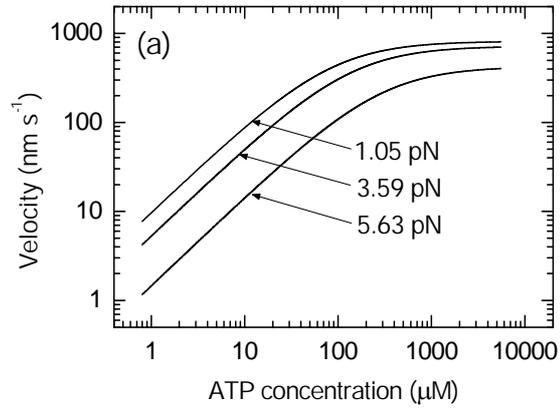

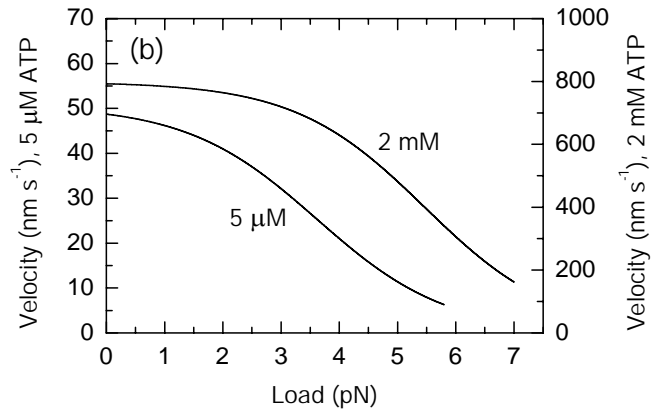

Fig. 3

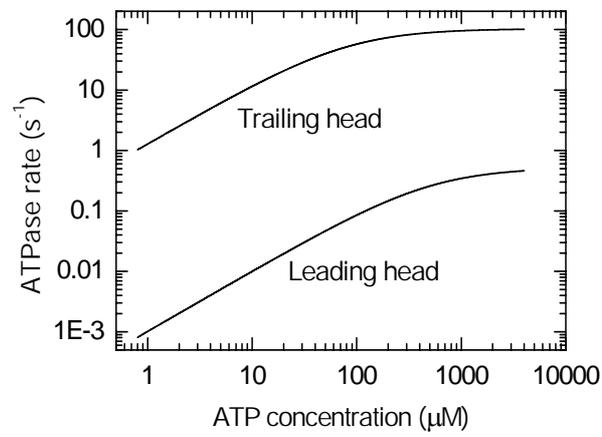

Fig. 4

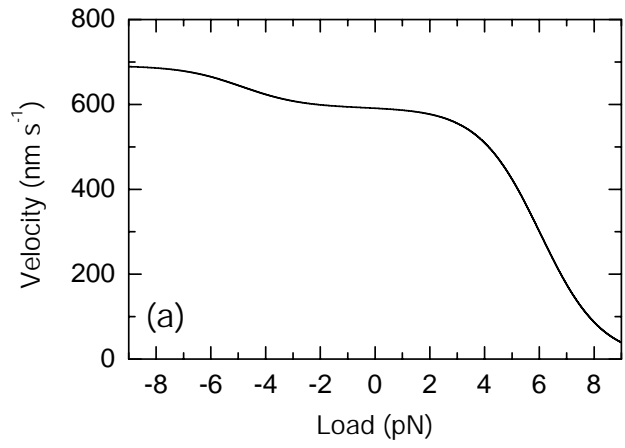

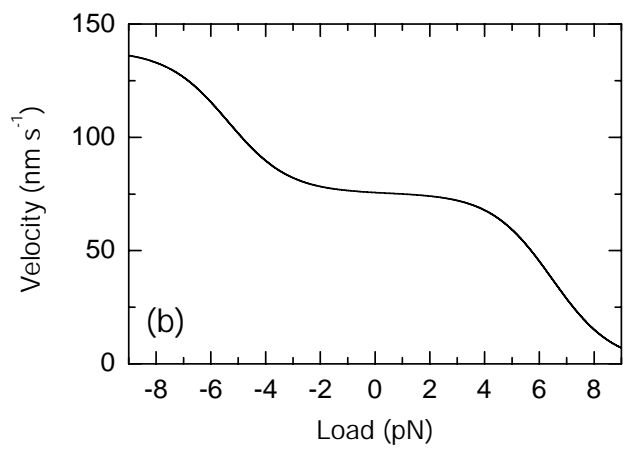

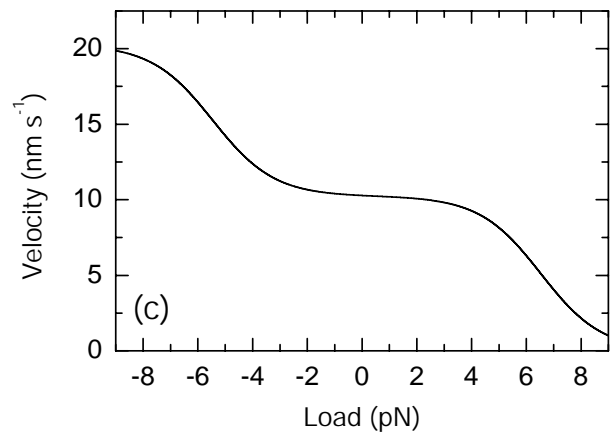

Fig. 5

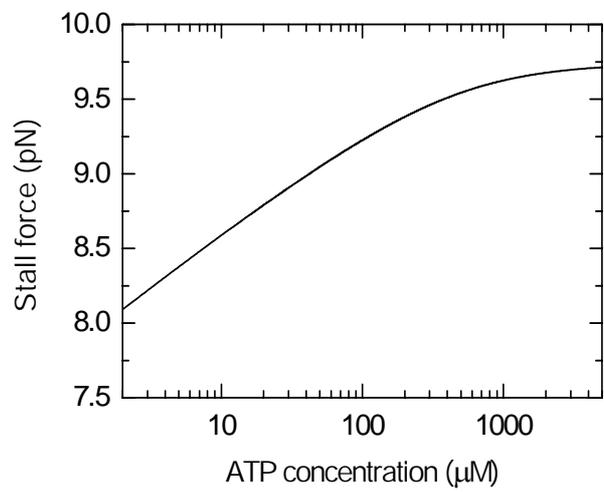

Fig. 6